\documentstyle[amssymb,12pt]{article}

\begin{document}

\title{Some solutions of the Gauss-Bonnet Gravity with Scalar Field in Four Dimensions}
\author{ Metin G{\" u}rses\\
{\small Department of Mathematics, Faculty of Sciences}\\
{\small Bilkent University, 06800 Ankara - Turkey}}
\begin{titlepage}
\maketitle
\begin{abstract}
We give all  exact solutions of the Einstein-Gauss-Bonnet Field
Equations coupled with a scalar field in four dimensions under
certain assumptions. The main assumption we make in this work is
to take the second covariant derivative of the coupling function
proportional to the spacetime metric tensor. Although this
assumption simplifies the field equations considerably, to obtain
exact solutions we assume also that the  spacetime metric is
conformally flat.  Then we obtain a class of exact solutions.
\end{abstract}
\end{titlepage}

Recently there is an increasing interest in the Gauss-Bonnet
theory with a scalar field to  look for possible theoretical
explanation to some cosmological problems such as acceleration of
the universe \cite{noj1}.  Accelerated cosmological solutions were
first suggested in \cite{odin1}, \cite{odin2} and also discussed
in \cite{mota1}, \cite{mota2}. It is also expected that this
theory or its modifications may have some contributions to some
astrophysical phenomena. For this purpose, spherically symmetric
solutions of this theory were first studied in \cite{des1},
\cite{des2}. It has been observed that the Post-Newtonian
approximation does not give any new contribution in addition to
the post-Newtonian parameters of the general relativity
\cite{teb}. Black hole solutions in the framework of the GB
gravity are investigated recently  in \cite{galt} (see also
\cite{mig}, \cite{kant}). There are also attempts to find exact
solutions and to study the stability of the Gauss-Bonnet theory in
various dimensions with actions containing higher derivative
scalar field couplings \cite{pcd}, \cite{jak}, \cite{dav}.

Since the Gauss-Bonnet term is a topological invariant in four
dimensions  it does not contribute to the Einstein field
equations. On the other hand it contributes to the field equations
if it couples to a spin-0 zero field.  In this work we consider a
four dimensional action containing the Einstein-Hilbert part,
massless scalar field and the Gauss-Bonnet term coupled with the
scalar field.  The corresponding action is given by \cite{teb}

\begin{equation}
S=\int d^4x\, \sqrt{-g}\, [ {R \over 2\kappa^2}-{1 \over 2}
\partial_{\mu} \phi\, \partial^{\mu}\, \phi-V(\phi)+f(\phi) GB]
\end{equation}
where $\kappa^2=8\pi G$ ($c=\hbar=1$) and

\begin{equation}\label{den00}
GB=R^2-4R^{\alpha \beta} R_{\alpha \beta}+R^{\alpha \beta \sigma
\gamma} R_{\alpha \beta \sigma \gamma}
\end{equation}
and $f$ is an arbitrary function of the scalar filed $\phi$
(coupling function). Here $V$ is potential term for the scalar
field. The field equations are given by

\begin{eqnarray}\label{den0}
R_{\mu \nu}&=&\kappa^2\, [{1 \over 2} \partial_{\mu}\, \phi
\partial_{\nu}\, \phi +{1 \over 2} V(\phi)\, g_{\mu \nu}+2
(\nabla_{\mu} \nabla_{\nu}f)R-g_{\mu \nu}\, (\nabla^{\rho}\nabla_{\rho} f) R \nonumber \\
&&-4(\nabla^\rho \nabla_{\mu}f )R_{\nu \rho} -4(\nabla^\rho
\nabla_{\nu}f )R_{\mu \rho}+4 (\nabla^{\rho}\nabla_{\rho} f) R_{\mu \nu} \nonumber \\
&&+2g_{\mu \nu} (\nabla^{\rho} \nabla^{\sigma} f)R_{\rho \sigma}-4
(\nabla^{\rho} \nabla^{\sigma} f) R_{\mu \rho \nu \sigma}]
\end{eqnarray}

\begin{equation}\label{den2}
\nabla^{\rho} \nabla_{\rho} \phi-V^{\prime}(\phi)+f^{\prime} GB=0
\end{equation}

Einstein field equations are usually solved under certain
assumptions like spherical symmetry, plane symmetry and axial
symmetry. In some cases we assume a form for the spacetime metric
like conformally flat, Kerr-Schild and G{\" o}del types. In each
one we create a class of exact solutions of Einstein's field
equations \cite{cramer}. In this work our intention is open such a
direction in GB theory and obtain exact solutions of this theory
and its modifications under certain assumptions. To this end we
now assume the spacetime geometry $(M,g)$ is such that ({\it
assumption 1})

\begin{equation} \label{den1}
\nabla_{\mu} \nabla_{\nu} f=\Lambda_{1} g_{\mu \nu}+\Lambda_{2}
\ell_{\mu} \ell_{\nu}
\end{equation}
where $\Lambda_{1}$ and $\Lambda_{2}$ are scalar functions and
$\ell_{\mu}$ is a vector field. In the sequel we will assume that
$\Lambda_{2}=0$ ({\it assumption 2}). Eq.(\ref{den1}) restricts
the space-time $(M,g)$. Among these space-times admitting
(\ref{den1}) we have conformally flat space-times ({\it assumption
3}).

\begin{equation}\label{met}
g_{\mu \nu}=\psi^{-2}\, \eta_{\mu \nu}
\end{equation}
where $\psi$ is a scalar function. In such space-times the
conformal tensor vanishes identically. Hence

\begin{equation}
GB=-2R^{\alpha \beta} R_{\alpha \beta}+{2 \over 3}R^2
\end{equation}
Then the field equations (\ref{den0}) reduce to

\begin{equation}\label{den3}
(1-4\Lambda_{1} \kappa^2)\,R_{\mu \nu}=\kappa^2\, [{1 \over 2}
\partial_{\mu}\, \phi
\partial_{\nu}\, \phi +{1 \over 2} V(\phi)\, g_{\mu \nu}]
\end{equation}
We have now the last assumption: All functions depend on
$z=k_{\mu}x^{\mu}$ where $k_{\mu}$ is a constant vector,
$\partial_{\mu} k_{\nu}=0$. Then from (\ref{den1}) we get

\begin{equation}
f^{\prime}=C \psi^{-2}, \Lambda_{1}=-Ck^2 {\psi^{\prime}\over
\psi}
\end{equation}
where $C$ is an arbitrary constant and $k^2 =\eta^{\mu \nu}k_{\mu}
k_{\nu}$. By using (\ref{den3}) and the Ricci tensor

\begin{equation}\label{ric}
R_{\mu \nu}=2{\psi_{,\mu \nu} \over \psi}+[{1 \over \psi}
\eta^{\alpha \beta}\, \psi_{,\alpha \beta}-{3 \over \psi^2}\,
\eta^{\alpha \beta}\, \psi_{,\alpha} \psi_{,\beta}]\, \eta_{\mu
\nu},
\end{equation}
for the metric (\ref{met}) we obtain the following  equations

\begin{eqnarray}
(1-4\Lambda_{1}\kappa^2)\psi^{-1} \psi^{\prime \prime}={\kappa^2
\over 4}
(\phi^{\prime})^2,\label{den4}\\
V=-{2k^2 \over
\kappa^2}(1-4\Lambda_{1}\kappa^2)[3(\psi^{\prime})^2-\psi
\psi^{\prime \prime}],\label{pot}\\
k^2 \psi^4 (\psi^{-2} \phi^{\prime})^{\prime}-\dot{V}+\dot{f}
GB=0, \label{den5}\\
f^{\prime}=C \psi^{-2}, \Lambda_{1}=-Ck^2 {\psi^{\prime} \over
\psi}
\end{eqnarray}
where
\begin{equation}
GB=72(k^2)^2\psi^4[(\psi^{-1} \psi^{\prime})^2-\psi^{-1}
\psi^{\prime \prime}](\psi^{-1} \psi^{\prime})^2
\end{equation}
and a dot over a letter denotes derivative with respect to the
scalar field $\phi$.  Eqs(\ref{den4}) and (\ref{den5}) give
coupled  ODEs for the functions $\psi$ and $\phi$. Letting
$\psi^{\prime}/\psi=u$ and $\phi^{\prime}=v$ then these equations
become

\begin{eqnarray}
(1+4Ck^2 \kappa^2 u)(u^{\prime}+u^2)={\kappa^2 \over 4} v^2, \label{eq01}\\
 k^2\, \psi^2\, [(v^{\prime}-2uv)v-27Ck^2\,u^{\prime}u^2]=
 V^{\prime} \label{eq02}
 \end{eqnarray}
where $V$ is given by (from (\ref{pot}))

\begin{equation}
V=-{2k^2 \over \kappa^2} (1+4Ck^2 \kappa^2 u)(2u^2-u^{\prime})\,
\psi^2 \label{eq03}
\end{equation}
Inserting $V$ from (\ref{eq03}) into Eq. (\ref{eq02}) (and using
(\ref{eq01}) in (\ref{eq02})) we obtain simply

\begin{equation}
3C (k^2)^2 u^2 u^{\prime}=0
\end{equation}

\noindent
 Hence we have the following solutions.

\vspace{0.3cm}

{\bf (A)\,\, $C=0$:} This corresponds to pure Einstein field
equations with a massless scalar field. The effect of the Gauss
Bonnet term disappears. Solutions of these field equations have
been given in \cite{gur}

\vspace{0.3cm}

 {\bf (B)\,\, $k^2=0$:} The vector field $k_{\mu}$ is null. Then
the only field equation is

\begin{equation}\label{denk14}
u^{\prime}+u^2={\kappa^2 \over 4} v^2
\end{equation}
and $V$ becomes zero. There is a single equation for the two
fields $u$ and $v$. This means that, if one of the fields $u$ or
$v$ is given then the other one is determined directly. The metric
takes the form

\begin{equation}
ds^2=\psi(p)^{-2}\, [2dp dq+dx^2+dy^2]
\end{equation}
where $p$ and $q$ are null coordinates and
$k_{\mu}=\delta_{\mu}^{p}$ and the above equation (\ref{denk14})
becomes

\begin{equation}
\psi_{pp}={\kappa^2 \over 4} (\phi^{\prime})^2\, \psi
\end{equation}
and the Einstein tensor represents a null fluid with zero
pressure.

\begin{equation}
G_{\mu \nu}={\kappa^2 \over 2}\, (\phi^{\prime})^2\, k_{\mu}\,
k_{\nu}
\end{equation}
Although the coupling function $f$ is nonzero the effect of the GB
term is absent in this type. Such a class of solutions belongs to
class (A).

 \vspace{0.3cm}

 {\bf (C) $k^2 \ne 0$:} The vector field $k_{\mu}$ is non-null.
 Then $u=m$ a real constant which leads to the following solution.

\begin{equation}
\psi=\psi_{0}\, e^{m\,z}, ~~\phi=\phi_{0}+\phi_{1}\,z
\end{equation}
where $\psi_{0}$ and $\phi_{0}$ are arbitrary constants and

\begin{equation}
(1+4Ck^2 \kappa^2\,m)m^2\, ={\kappa^2 \over 4}\, \phi_{1}^2,~~V
=-k^2\, \phi_{1}^2\, \psi^2
\end{equation}
where $\phi_{1} \ne 0$. The potential function $V$ takes the form

\begin{equation}
V(\phi)= V_{0}\, e^{\pm \, {\phi \over \xi}}, ~~V_{0}=-k^2\,
\phi_{1}^2\, \psi_{0}^2\, e^{\mp  {\phi_{0} \over \xi}}
\end{equation}
where $\xi=1+4Ck^2\, \kappa^2\, m$ and coupling function $f$ takes
the form

\begin{equation}
f=f_{0}-f_{1}\, e^{\mp {\phi \over \xi}},~~~ f_{1}=(C / \xi\,
\psi_{0}^2)\,e^{\mp \phi_{0} \over \xi}
\end{equation}
The solution we obtained here is free of singularities but not
asymptotically flat. On the other hand, by using this solution it
is possible to obtain an asymptotically flat cosmological
solution.

\vspace{0.3cm}

This solution is well understood in a new  coordinate chart
$\{x^a,t\}$ where the line element takes the following form (after
a scaling)

\begin{equation}
ds^2={t^2 \over t_{0}^2}\, \eta_{ab}\,dx^a dx^b+\epsilon dt^2
\end{equation}
where $t_{0}$ is a nonzero  constant. If $t$ is a spacelike
coordinate then $\epsilon=1$ and Latin indices take values
$a=0,1,2$. If $t$ is a timelike coordinate then $\epsilon=-1$ and
Latin indices take values $a=1,2,3$. $\eta_{ab}$ is the  metric of
the flat three dimensional geometry orthogonal to the
$u$-direction. The Ricci tensor of the four dimensional metric

\begin{equation}
R_{tt}=0,~~ R_{ta}=0,~~ R_{ab}=-{2\epsilon \over
t_{0}^2}\,\eta_{ab}
\end{equation}
Hence the solution takes the form

\begin{equation}
\phi^{\prime}=\pm {2\sqrt{\xi} \over t}, ~~V(\phi)=-{4\epsilon \xi
\over t^2}
\end{equation}
where $\xi=1+4\kappa^2 C$,~ $\Lambda_{1}=C$ a constant, and
$f=f_{0}+{\epsilon C \over 2} t^2$ , $f_{0}$ is an arbitrary
constant. The curvature scalars are given by

\begin{equation}
R={6 \over t^2},~~ R_{\mu \nu}\,R^{\mu \nu}={12 \over t^4}
\end{equation}
and the Gauss-Bonnet scalar density $GB=0$. It clear that $t=0$ is
the spacetime singularity. Letting
$u_{\alpha}=\delta_{\alpha}^{t}$, the Einstein tensor becomes

\begin{equation}
G_{\alpha \beta}={2 \over t^2}\, u_{\alpha}\, u_{\beta}+{\epsilon
\over t^2}\, g_{\alpha \beta}
\end{equation}
This tensor has a physical meaning when $\epsilon=-1$ in which
case the Gauss-Bonnet gravity produces a singular cosmological
model. The Einstein tensor represents a perfect fluid with an
energy density $\rho=3/t^2$ and a negative pressure $p=-1/t^2$.
Both of them are singular at $t=0$.

\vspace{0.3cm}

We have found the most general solutions of the Gauss-Bonnet
gravity coupled to a scalar field under the assumptions stated in
the text. One solution  (B) depends on a null coordinate whose
Einstein tensor corresponds to the energy momentum tensor of a
null fluid with zero pressure. The other solution (C) depends on
variable $t$ whose curvature invariants are all singular at $t=0$.
When $t$ represents the time coordinate then GB gravity gives a
cosmological model with a negative pressure. The solution is
singular on the 3-surface $t=0$.

We would like to conclude with a remark. The field equations
(\ref{den0}) and (\ref{den2}) of the GB theory with a scalar field
resemble to the field equations of the modified Gauss-Bonnet
theory \cite{noj1}, \cite{noj2}. In the latter case the scalar
field $\phi$ and the potential term $V(\phi)$ are absent in the
action and the function $f=f(GB)$ depends on the GB term
(\ref{den00}). We remark that the flat metric is the only solution
of the modified Gauss-Bonnet field equations under the assumptions
made in the text. It seems that scalar field is crucial to obtain
non-flat metrics. It is however interesting to search for the
solutions of the modified GB field equations. For this purpose we
are planning to relax our assumptions 2 and 3 in a forthcoming
publication.

\vspace{0.4cm}

 I would like to thank the referees for their
constructive comments. This work is partially supported by the
Scientific and Technological Research Council of Turkey (TUBITAK)
and Turkish Academy of Sciences (TUBA).

\end{document}